\documentclass[reprint,superscriptaddress,showpacs,amsmath,amssymb,pra,aps]{revtex4-1}

\usepackage{amsmath}
\usepackage{amssymb}
\usepackage{graphics}
\usepackage{graphicx}% Include figure files
\usepackage{dcolumn}% Align table columns on decimal point
\usepackage{txfonts}% bold math
\usepackage[pdfpagemode=UseNone,pdfstartview=FitH,colorlinks=true,linkcolor=blue,urlcolor=blue,anchorcolor=blue,citecolor=blue]{hyperref}
\usepackage{ulem}
\allowdisplaybreaks[4]

\begin{document}

\title{Experimentally accessible quantum phase transition in a non-Hermitian Tavis-Cummings model engineered with two drive fields}

\author{Guo-Qiang Zhang}
\affiliation{Interdisciplinary Center of Quantum Information, State Key Laboratory of Modern Optical Instrumentation, and Zhejiang Province Key Laboratory of Quantum Technology and Device, Department of Physics, Zhejiang University, Hangzhou 310027, China}

\author{Zhen Chen}
\affiliation{Interdisciplinary Center of Quantum Information, State Key Laboratory of Modern Optical Instrumentation, and Zhejiang Province Key Laboratory of Quantum Technology and Device, Department of Physics, Zhejiang University, Hangzhou 310027, China}
\affiliation{Quantum Physics and Quantum Information Division, Beijing Computational Science Research Center, Beijing 100193, China}

\author{J. Q. You}
\email{jqyou@zju.edu.cn}
\affiliation{Interdisciplinary Center of Quantum Information, State Key Laboratory of Modern Optical Instrumentation, and Zhejiang Province Key Laboratory of Quantum Technology and Device, Department of Physics, Zhejiang University, Hangzhou 310027, China}

\date{\today}

\begin{abstract}
We study the quantum phase transition (QPT) in a non-Hermitian Tavis-Cummings (TC) model of experimentally accessible parameters, which is engineered with two drive fields applied to an ensemble of two-level systems (TLSs) and a cavity, respectively. When the two drive fields satisfy a given parameter-matching condition, the coupled cavity-TLS ensemble system can be described by an effective standard TC Hamiltonian in the rotating frame. In this ideal Hermitian case, the engineered TC model can exhibit the super-radiant QPT with spin conservation at an experimentally-accessible critical coupling strength, but the QPT is, however, spoiled by the decoherence. We find that in this non-Hermitian case, the QPT can be recovered by introducing a gain in the cavity to balance the loss of the TLS ensemble. Also, the spin-conservation law is found to be violated due to the decoherence of the system. Our study offers an experimentally realizable approach to implementing QPT in the non-Hermitian TC model.
\end{abstract}

\maketitle

\section{Introduction}

Quantum phase transition (QPT) has been widely studied in various quantum systems (see, e.g.,~\cite{Mebrahtu12,Hwang15,Hwang16,Zhou16,Wang14,Aharony96,Sachdev93,Harris08,Zhang13,Zhang17,Lv18,Felicetti20,He15}), because of its fundamental importance in quantum physics and potential applications in quantum technologies~\cite{Sachdev11}. Among them, the super-radiant QPT was predicted~\cite{Hepp73,Wang73,Hioe73,Hepp73-1} in, e.g., the Dicke model~\cite{Dicke54}, which involves the collective interaction between an ensemble of two-level systems (TLSs) and the quantized field in a cavity. In the thermodynamic limit of a large number of TLSs, the ground-state properties of the Dicke model, such as the excitations in the TLS ensemble, can display an abrupt change related to the QPT in the system~\cite{Emary03,Emary03-1,Ye11}, when continuously varying the collective coupling strength around a critical value. As required to be comparable to the frequencies of the TLS ensemble and the cavity mode, this critical coupling strength is difficult to achieve experimentally. Moreover, the no-go theorem due to the squared electromagnetic vector potential also hinders the presence of the super-radiant QPT in the Dicke model~\cite{Rzazewski75,Birula79,Rabl16}. Due to these difficulties, the nonequilibrium QPT (i.e., the simulated QPT) has been proposed~\cite{Dimer07,Zou14,Nagy10} and observed~\cite{Baumann10,Baumann11,Baden14} in cavity quantum electrodynamics (QED) systems by engineering an effective Dicke Hamiltonian.

In the experiment, most quantum systems can only reach the strong-coupling regime~\cite{Xiang13,Kurizki15}, where the Dicke model can be reasonably reduced to the Tavis-Cummings (TC) model~\cite{Tavis68} via the rotating-wave approximation (RWA)~\cite{Walls94}, i.e., ignoring the counter-rotating coupling terms between the TLS ensemble and the cavity mode. Similar to the Dicke model, the TC model is another archetypal model known to exhibit the super-radiant QPT~\cite{Castanos09}. Nevertheless, QPT in the TC model can occur only in the regime of extremely strong coupling, which is unaccessible for a realistic system and in which the RWA actually breaks down~\cite{Walls94}. In addition, the decoherence inevitably occurring in the realistic system can spoil the QPT in the TC model~\cite{Keeling10,Larson17,Soriente18,Feng15}. Therefore, it is of great importance to explore the nonequilibrium QPT by engineering an effective experimentally accessible TC Hamiltonian~\cite{Feng15,Zou13,Zhu20}.

In this paper, we study the QPT in a driven quantum system consisting of a TLS ensemble coupled to a cavity mode. To reduce the critical coupling strength of the TC model, we use two external fields with the same frequency to drive the TLS ensemble and the cavity, respectively. By choosing proper drive-field parameters, an effective TC model can be rebuilt in the rotating frame, with the effective frequencies of the TLS ensemble and the cavity mode being their respective frequency detunings from the drive fields. In the ideal {\it Hermitian} case (i.e., without decoherence in the system), we can demonstrate the QPT in the driven system, where the corresponding critical coupling strength is tunable (via varying the drive-field frequency) and experimentally accessible. However, decoherence inevitably occurs in the system and the QPT of the TC model is spoiled in this {\it non-Hermitian} case~\cite{Keeling10,Larson17,Soriente18,Feng15}. We find that the QPT in the non-Hermitian TC model can be recovered by using a cavity with gain to {\it balance} the loss of the TLS ensemble. In sharp contrast to the spin conservation in the ideal Hermitian case, the spin-conservation law is found to be violated in the non-Hermitian case~\cite{Torre16,Kirton17}. Moreover, we propose to implement this non-Hermitian TC model with a hybrid circuit-QED system.

As indicated above, we propose to engineer an experimentally accessible {\it non-Hermitian} TC model that can have QPT in the presence of decoherence in the system.
By engineering cavity-mediated Raman transitions, an effective Dicke Hamiltonian can be built in a cavity QED system consisting of an ensemble of four-level systems~\cite{Dimer07,Baden14} or three-level systems~\cite{Zou14} coupled to a quantized cavity field. In these schemes~\cite{Dimer07,Zou14,Baden14}, the two drive fields both directly pump the ensemble and the nonequilibrium QPT in the effective Dicke model is demonstrated by tuning the frequencies and magnitudes of the two drive fields.
In addition, the spatial self-organization of a Bose-Einstein condensate (BEC) inside an optical cavity can be mapped to the Dicke Hamiltonian~\cite{Nagy10,Baumann10,Baumann11}, where the motional degree of freedom of the BEC is coupled to the cavity mode and the BEC is driven by a far-detuned laser field. On the contrary, in our proposal, only an ensemble of TLSs, rather than the ensemble of multilevel systems or the BEC, are used and the two drive fields pump the TLS ensemble and the cavity, respectively. This exploits a different mechanism to engineer the model.

\section{QPT in the ideal Hermitian case}

\begin{figure}[tbp]
\centering
\includegraphics[width=0.45\textwidth]{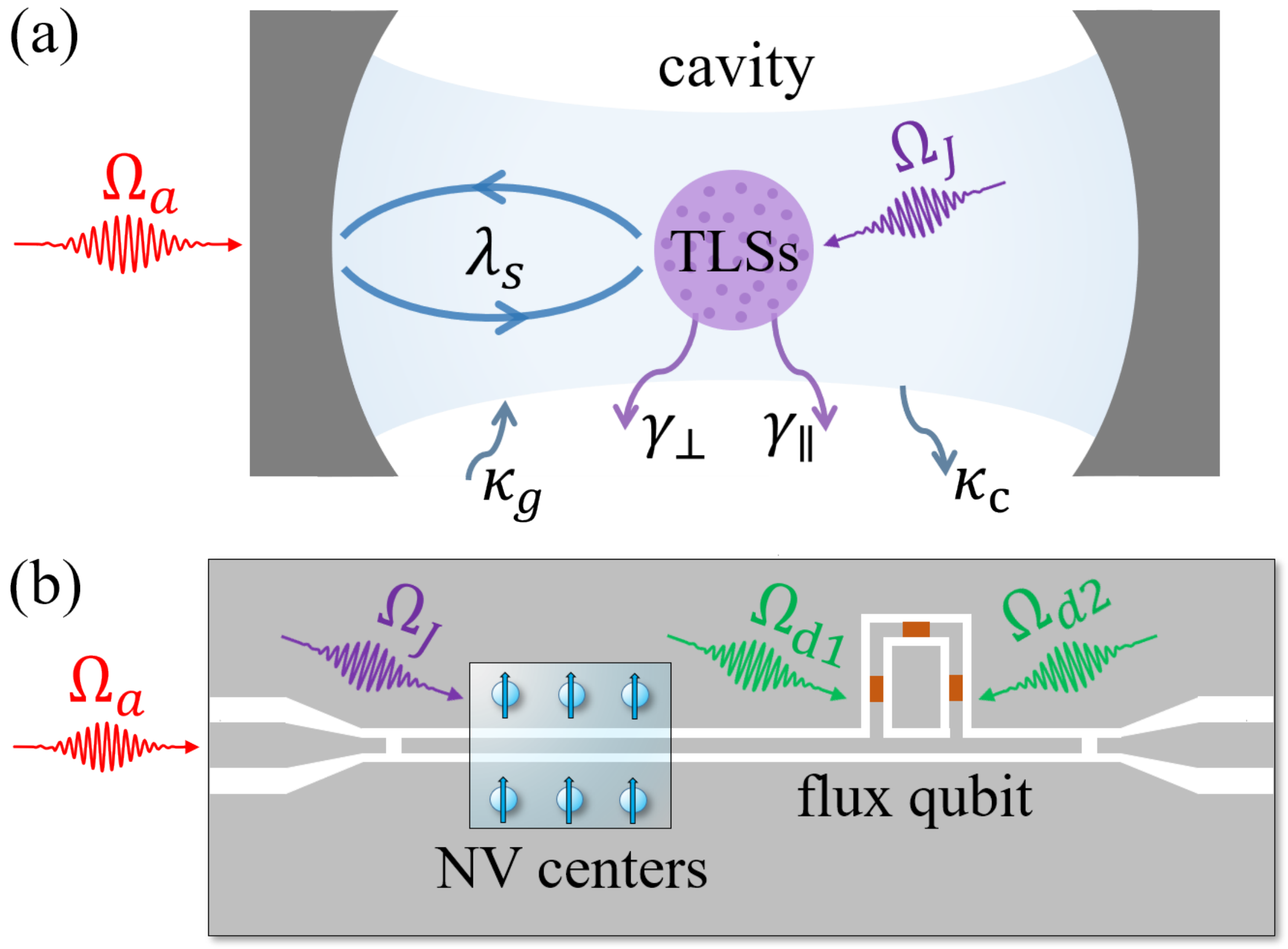}
\caption{(a) Schematic representation of the engineered TC model with a TLS ensemble strongly coupled to a cavity mode, where two drive fields with the same frequency are applied to the ensemble and the cavity mode, respectively. (b) Implementation of the proposed TC model using a hybrid circuit-QED system composed of an ensemble of NV centers in diamond interacting with an active coplanar-waveguide resonator, where the gain of the resonator results from an auxiliary flux qubit controlled by two microwave fields~\cite{Quijandria18}. Also, two fields with the same frequency drive the NV ensemble and the resonator, respectively.}
\label{figure1}
\end{figure}

The proposed system consists of $N$ TLSs (e.g., spins) in a cavity, each with the same transition frequency $\omega_{s}$ and coupled to a cavity mode with coupling strength $\lambda_{s}$ [see Fig.~\ref{figure1}(a)]. This system can be described in a RWA by the TC model (we set $\hbar=1$),
\begin{equation}\label{TC}
H_{\rm TC}=\omega_{c}a^{\dag}a+\omega_{s}J_{z}+\frac{\lambda}{\sqrt{N}}(a^{\dag}J_{-}+aJ_{+}),
\end{equation}
where $a$ ($a^{\dag})$ is the annihilation (creation) operator of the cavity mode with resonant frequency $\omega_c$, $J_{x}, J_{y}$, and $J_{z}$ are the collective spin operators of the TLS ensemble with raising and lowering operators $J_{\pm}=J_{x} \pm iJ_{y}$, and $\lambda=\lambda_{s}\sqrt{N}$ is the collective coupling strength between the TLS ensemble and the cavity mode.
This Hamiltonian has a conserved parity~\cite{Castanos09}, $[H_{\rm TC},\Pi]=0$, where $\Pi=\exp{[i\pi(a^{\dag}a+J_z+N/2)]}$.
In the ideal case without decoherence in the system, the TC model exhibits a QPT at the critical coupling strength $\lambda = \lambda_{c} \equiv \sqrt{\omega_{s}\omega_{c}}$ in the thermodynamic limit $N \rightarrow +\infty$~\cite{Castanos09}. However, this QPT is spoiled by the decoherence of the system~\cite{Keeling10,Soriente18,Larson17,Feng15}. Also, it is extremely difficult to demonstrate the QPT due to the inaccessibility of the very large critical coupling strength $\lambda_{c}$ in a realistic system.

To solve these problems, we first manage to reduce the critical coupling strength by applying two drive fields of the same frequency $\omega_{d}$ to the cavity and the TLS ensemble, respectively. This corresponds to adding the drive Hamiltonian $H_d=(\Omega_{a}a^{\dag}e^{-i\omega_d t}+\Omega_{a}^{*}ae^{i\omega_d t}) +(1/\sqrt{N})(\Omega_{J}J_{+}e^{-i\omega_d t}+\Omega_{J}^{*}J_{-}e^{i\omega_d t})$ to the Hamiltonian in Eq.~(\ref{TC}). Here $\Omega_{a}$ is the Rabi frequency between the drive field and the cavity mode, and $\Omega_{J}\equiv\Omega_{s}\sqrt{N}$ is the collective Rabi frequency between the drive field and the TLS ensemble, where $\Omega_{s}$ is the Rabi frequency between the drive field and each TLS. In a rotating reference frame with respect to the frequency $\omega_{d}$ of the two drive fields, the total Hamiltonian of the system can be converted to
\begin{eqnarray}\label{TC-d}
H^{(d)}_{\rm TC}&=&\Delta_{c} a^{\dag}a + \Delta_{s}J_{z}+\frac{\lambda}{\sqrt{N}}(aJ_{+}+a^{\dag}J_{-})
                    +(\Omega_{a}a^{\dag}+\Omega_{a}^{*}a) \nonumber\\
                & &+\frac{1}{\sqrt{N}}(\Omega_{J}J_{+}+\Omega_{J}^{*}J_{-}),
\end{eqnarray}
where $\Delta_{c(s)} \equiv \omega_{c(s)}-\omega_{d}$ ($>0$) is the frequency detuning of the cavity mode (TLS ensemble) relative to the corresponding drive field. By introducing a displacement $a= A+\alpha$ and  $a^{\dag}= A^{\dag}+\alpha^{*}$ with $\alpha=-\Omega_{a}/\Delta_{c}$, i.e., a translation transform, the above Hamiltonian becomes
\begin{eqnarray}\label{TC-u}
H_{\rm TC}^{(d)}&=&\Delta_{c} A^{\dag}A + \Delta_{s}J_{z}+\frac{\lambda}{\sqrt{N}}(AJ_{+}+A^{\dag}J_{-}) \nonumber\\
           & & +\frac{1}{\sqrt{N}}\left[(\Omega_{J}+\lambda\alpha)J_{+}+(\Omega_{J}^{*}+\lambda\alpha^{*})J_{-}\right],
\end{eqnarray}
where $A^{\dag}$ and $A$ are also bosonic creation and annihilation operators obeying $[A,A^{\dag}]=1$.
When the two drive fields satisfy the parameter-matching condition $\Omega_{a}/\Omega_{J}=\Delta_{c}/\lambda$, the effects of these two drive fields cancel each other and the Hamiltonian is reduced to
\begin{equation}\label{eff}
H_{\rm TC}^{(d)}=\Delta_{c} A^{\dag}A + \Delta_{s}J_{z}+\frac{\lambda}{\sqrt{N}}(AJ_{+}+A^{\dag}J_{-}),
\end{equation}
which has the same form as the standard TC model in Eq.~(\ref{TC}), but the critical coupling strength $\lambda = \lambda_{c} \equiv \sqrt{\Delta_{s}\Delta_{c}}$ becomes experimentally accessible by reducing $\Delta_c$ ($\Delta_s$) via tuning the drive-field frequency $\omega_d$. Like the standard TC model in Eq.~(\ref{TC}), the effective TC model in Eq.~(\ref{eff}) also exhibits a normal (super-radiant) phase with unbroken (broken) parity symmetry when $\lambda<\lambda_{c}$ ($\lambda>\lambda_{c}$), and the TLS ensemble in the thermodynamic limit has the critical behaviors~\cite{Zou13},
\begin{equation}\label{order-Jz}
\begin{split}
\frac{\langle J_{z}\rangle}{(N/2)}
=\Bigg\{
\begin{array}{cc}
                 -1,                         &~~~ \lambda < \lambda_{c};\\
  -\lambda_{c}^{2}/\lambda^{2},    &~~~ \lambda \geq \lambda_{c},
\end{array}
\end{split}
\end{equation}
and
\begin{equation}\label{order-J-}
\begin{split}
\frac{\langle J_{-}\rangle}{(N/2)}
=\Bigg\{
\begin{array}{cc}
                 0,                         &~~~ \lambda < \lambda_{c};\\
  (1-\lambda_{c}^{4}/\lambda^{4})^{1/2},    &~~~ \lambda \geq \lambda_{c}.
\end{array}
\end{split}
\end{equation}
This gives $\langle J_{z}\rangle \sim |\lambda-\lambda_{c}|^{\nu_{z}}$ and $\langle J_{-}\rangle \sim |\lambda-\lambda_{c}|^{\nu_{-}}$ around $\lambda=\lambda_{c}$, with critical exponents $\nu_{z}=1$ and $\nu_{-}=1/2$. The displaced cavity field can also display a QPT because $\langle A \rangle=-\lambda\langle J_{-}\rangle/(\sqrt{N}\Delta_{c})$, i.e., $\langle A^{\dag}A \rangle$ behaves as $\langle A^{\dag}A \rangle\sim |\lambda-\lambda_c|^{\nu_{A}}$, with $\nu_{A}=1$.  Therefore, $\langle a^{\dag}a \rangle\sim|\lambda-\lambda_c|^{\nu_a}$, with $\nu_{a}=1/2$ (see Appendix \ref{appendix-A}), owing to $\langle a \rangle \equiv \langle A \rangle+\alpha=-\lambda\langle J_{-}\rangle/(\sqrt{N}\Delta_{c})-\Omega_{a}/\Delta_{c}$. Note that two suitable drive fields are required to demonstrate the QPT. If only one finite drive field is applied (i.e., either $\Omega_{a}\neq 0$ but $\Omega_{J}=0$ or $\Omega_{J} \neq 0$ but $\Omega_{a} = 0$), the effective Hamiltonian of the driven system does not preserve the parity symmetry~\cite{Emary04}. In this case, the system can only tend to exhibit the critical behavior when the sole drive field becomes extremely weak~\cite{Zou13}.

\section{QPT in the non-Hermitian case with gain}\label{non-Hermitian}

\subsection{The passive-cavity case with loss}

When the decoherence of the system is considered, the system becomes non-Hermitian.
For the standard TC model, as demonstrated in Refs.~\cite{Soriente18,Larson17}, the QPT is spoiled by the presence of any decay of the cavity. When including the damping of the TLS ensemble, this limitation still exists [see Eq.~(\ref{spin-1}) below]. This discontinuity looks {\it counterintuitive}, but is actually an {\it intrinsic} characteristic of the TC model. In most cases when considering the damping of the system, the Dicke model can exhibit the QPT~\cite{Torre16,Nagy11,Brennecke13}, but it also has the similar characteristic under specific circumstances. For instance, when the decay of the cavity is considered but the radiative decay of the TLS ensemble is ignored, the presence of an {\it infinitesimal} nonradiative dephasing of the TLS ensemble can completely spoil the QPT in the Dicke model~\cite{Kirton17}. Below we first show that in our engineered TC model, the spoiling of the QPT by an infinitesimal damping of the system occurs only when the parameter-matching condition for the two drive fields is exactly satisfied. If this condition is loosened, the system can have nontrivial solutions in the presence of the decoherence, but the corresponding behaviors deviate from the QPT of the system. Then, we study how to recover the QPT by introducing a gain medium in the cavity.

Here we use a quantum Langevin approach~\cite{Walls94} to study the critical behavior of the non-Hermitian system.
With the Hamiltonian in Eq.~(\ref{TC-d}), the dynamics of the driven system is governed by the following Langevin equations:
\begin{eqnarray}\label{Langevin}
\dot{a}&=&-i(\Delta_{c}-i\kappa_{c})a-i\frac{\lambda}{\sqrt{N}}J_{-}-i\Omega_{a}
           +\sqrt{2\kappa_{c}}\,a_{\rm in}, \nonumber\\
\dot{J_{-}}&=&-i(\Delta_{s}-i\gamma_{\perp})J_{-}+i\frac{2\lambda}{\sqrt{N}}J_{z}a
                      +i2J_{z}\frac{\Omega_{J}}{\sqrt{N}}+\sqrt{2\gamma_{\bot}}\,J_{\rm -,\, in},\nonumber\\
\dot{J_{z}}&=&-i\frac{\lambda}{\sqrt{N}}(aJ_{+}-a^{\dag}J_{-})-i\frac{1}{\sqrt{N}}(\Omega_{J}J_{+}-\Omega_{J}^{*}J_{-})\\
           & & -\gamma_{\parallel}\bigg(\frac{N}{2}+J_{z}\bigg)
               +\sqrt{2\gamma_{\parallel}}\,J_{\rm z,\, in},\nonumber
\end{eqnarray}
where $\kappa_{c}$ is the decay rate of the cavity, $\gamma_{\perp}$ ($\gamma_{\parallel}$) is the transversal (longitudinal) relaxation rate of the TLS ensemble, and $a_{\rm in}$ as well as
$J_{\rm -,\, in}$ and $J_{\rm z,\, in}$ are the input noise operators related to the cavity and TLS ensemble, with $\langle a_{\rm in}\rangle=\langle J_{\rm -,\, in}\rangle=\langle J_{\rm z,\, in}\rangle=0$.
For any operator $\mathcal{O}$, it can be written as a sum of its expected value $\langle \mathcal{O} \rangle$ and fluctuation $\delta \mathcal{O}$, i.e., $\mathcal{O}=\langle \mathcal{O} \rangle+\delta \mathcal{O}$. It follows from Eq.~(\ref{Langevin}) that the expected values $\langle a \rangle$, $\langle J_{-} \rangle$ and $\langle J_{z} \rangle$ satisfy
\begin{equation}\label{mean}
\begin{split}
\langle\dot{a}\rangle=&-i(\Delta_{c}-i\kappa_{c})\langle a\rangle-i\frac{\lambda}{\sqrt{N}}\langle J_{-}\rangle-i\Omega_{a},\\
\langle\dot{J_{-}}\rangle=&-i(\Delta_{s}-i\gamma_{\perp})\langle J_{-}\rangle
                            +i2\frac{\lambda}{\sqrt{N}}\langle J_{z} a\rangle
                      +i2\langle J_{z}\rangle\frac{\Omega_{J}}{\sqrt{N}},\\
\langle\dot{J_{z}}\rangle=&-i\frac{\lambda}{\sqrt{N}}(\langle a J_{+}\rangle-\langle a^{\dag} J_{-}\rangle)
                               -i\frac{1}{\sqrt{N}}(\Omega_{J}\langle J_{+}\rangle-\Omega_{J}^{*}\langle J_{-}\rangle)\\
                          &       -\gamma_{\parallel}\bigg(\frac{N}{2}+\langle J_{z}\rangle\bigg).
\end{split}
\end{equation}
As shown in Appendix~\ref{appendix-B}, the above equations can also be derived using a master equation approach~\cite{Breuer07}.
At the steady state, $\langle\dot{a}\rangle=\langle\dot{J}_{-}\rangle=\langle\dot{J}_{z}\rangle=0$. From Eq.~(\ref{mean}), we then obtain
\begin{equation}\label{steady}
\begin{split}
&(\Delta_{c}-i\kappa_{c})\langle a\rangle+\frac{\lambda}{\sqrt{N}}\langle J_{-}\rangle+\Omega_{a}=0,\\
&(\Delta_{s}-i\gamma_{\perp})\langle J_{-}\rangle-\frac{2\lambda}{\sqrt{N}}\langle J_{z}\rangle \langle a \rangle
                      -2\langle J_{z}\rangle\frac{\Omega_{J}}{\sqrt{N}}=0,\\
&\frac{\lambda}{\sqrt{N}}(\langle a\rangle \langle J_{+}\rangle-\langle a^{\dag}\rangle \langle J_{-}\rangle)+\frac{1}{\sqrt{N}}(\Omega_{J}\langle J_{+}\rangle-\Omega_{J}^{*}\langle J_{-}\rangle)\\
&                   -i\gamma_{\parallel}\bigg(\frac{N}{2}+\langle J_{z}\rangle\bigg)=0,
\end{split}
\end{equation}
where a mean-field approximation is applied to the two-operator terms, i.e., $\langle J_{z} a \rangle=\langle J_{z} \rangle\langle a \rangle$ and $\langle aJ_{+}\rangle=\langle a\rangle\langle J_{+}\rangle$. Note that the mean-field approximation is usually used to study the QPT in both TC and Dicke models~\cite{Soriente18,Zhu20,Kirton17}, because it can give accurate results in the thermodynamic limit of the TLS ensemble (i.e., when the number of the TLSs is sufficiently large) (cf. the results and discussions in Ref.~\cite{Krimer19}). In Appendix~\ref{appendix-C}, we also show the results obtained beyond the mean-field approximation in Eq.~(\ref{steady}) and compare them with the results obtained using Eq.~(\ref{steady}). Actually, the difference between them is negligibly small.

Substituting the first equation in Eq.~(\ref{steady}) into the second and third equations in Eq.~(\ref{steady}) to eliminate $\langle a\rangle$, we have
\begin{equation}\label{spin-1}
\begin{split}
&\bigg[1+\frac{\lambda^{2}}{(\Delta_{c}\Delta_{s}-\kappa_{c}\gamma_{\perp})-i(\Delta_{c}\gamma_{\perp}+\Delta_{s}\kappa_{c})}
       \frac{\langle J_{z}\rangle}{(N/2)}\bigg]\langle J_{-}\rangle=0,\\
&\frac{\kappa_{c}\lambda^{2}}{\Delta_{c}^{2}+\kappa_{c}^{2}}\frac{\langle J_{+}\rangle \langle J_{-}\rangle}{(N/2)^{2}}
                                               +\gamma_{\parallel}\bigg(1+\frac{\langle J_{z}\rangle}{(N/2)}\bigg)=0,
\end{split}
\end{equation}
when using the parameter-matching condition for the two drive fields, $\Omega_{a}/\Omega_{J}=(\Delta_{c}-i\kappa_{c})/\lambda$, in the non-Hermitian case. Obviously, Eq.~(\ref{spin-1}) has only a set of {\it trivial} solutions $\langle J_{-}\rangle=0$ and $\langle J_{z}\rangle/(N/2)=-1$. This verifies that the decoherence of the system ruins the QPT occurring in the standard Hermitian TC model, which has also been demonstrated in Refs.~\cite{Soriente18,Larson17}.
When parameter mismatch occurs for the two drive fields, i.e., $\Omega_{a}/\Omega_{J} \neq (\Delta_{c}-i\kappa_{c})/\lambda$,
Eq.~(\ref{spin-1}) can have nontrivial solutions for $\langle J_{-}\rangle$ and $\langle J_{z}\rangle$, but the results considerably deviate from the cusp-like QPT behavior of the standard TC model [cf. Fig.~\ref{figure-s1}(b) in Appendix~\ref{appendix-C}], due to the decoherence of the system.

\begin{figure}[tbp]
\centering
\includegraphics[width=0.48\textwidth]{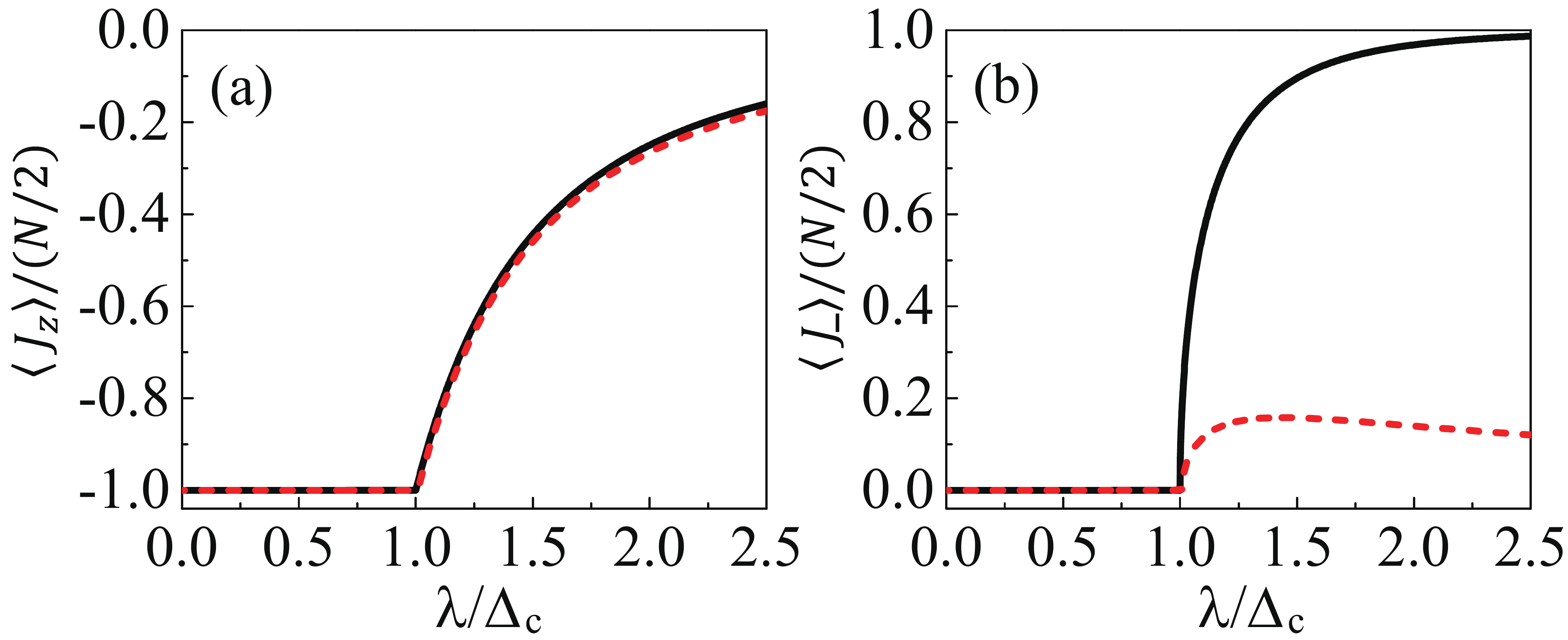}
\caption{(a) $\langle J_{z}\rangle/(N/2)$ and (b) $\langle J_{-}\rangle/(N/2)$ versus the reduced coupling strength $\lambda/\Delta_{c}$ under the parameter-matching condition of the two drive fields. For the (black) solid curve, $\kappa_{c}=\gamma_{\perp}=\gamma_{\parallel}=\kappa_{g}=0$, while $\kappa_{c}=\gamma_{\perp}=1$, $\gamma_{\parallel}=0.1$ and $\kappa_{g}=\kappa_{g}^{(0)} = \kappa_{c}+\gamma_{\perp}\Delta_{c}/\Delta_{s}$ for the (red) dashed curve. Other parameters are $\lambda=8$, $\Omega_{a}/\sqrt{N}=1$, and $\Delta_{s}=\Delta_{c}$.}
\label{figure2}
\end{figure}

\subsection{The active-cavity case with gain}

To recover the QPT, we introduce a gain medium in the dissipative cavity. With the gain included, we obtain the same equations as Eqs.~(\ref{Langevin})-(\ref{spin-1}), but $\kappa_{c}$ is replaced by $\kappa\equiv\kappa_{c}-\kappa_{g}$, where $\kappa_{g}$ is the gain rate of the cavity owing to the gain medium. Correspondingly, the parameter-matching condition for the two drive fields becomes $\Omega_{a}/\Omega_{J}=(\Delta_{c}-i\kappa)/\lambda$.
In the non-Hermitian system with parity-time symmetry (see, e.g.,~\cite{Peng14,Chang14}), a gain medium is also introduced in the cavity, where the damping rate $\kappa_{c}$ of the cavity is replaced by $\kappa_{c}-\kappa_{g}$ as well. It balances the loss and gain in the system to have real eigenvalues for the non-Hermitian Hamiltonian. In the non-Hermitian system we consider, Eq.~(\ref{spin-1}) has only a set of real but trivial solutions $\langle J_{-}\rangle=0$ and $\langle J_{z}\rangle/(N/2)=-1$. Also, Eq.~(\ref{spin-1}) has another set of solutions for $\langle J_{-}\rangle$ and $\langle J_{z}\rangle$, but this set of solutions is not physical because the obtained $\langle J_{z}\rangle$ is complex.
Here, a gain medium is also introduced in the cavity to balance the loss and gain in the system, but the purpose is to have the complex solution of $\langle J_{z}\rangle$ become real. As for the gain medium, it can be different for different types of cavities. For instance, driven rare-earth-metal ions are often used as the gain medium in optical cavities~\cite{Peng14,Chang14}. In Sec.~IIIC below, we will discuss the possible implementation of the hybrid system, where a driven flux qubit can be used as the gain medium of a coplanar-waveguide resonator~\cite{Quijandria18}.

Therefore, to recover the QPT, it is required that $\Delta_{c}\gamma_{\perp}+\Delta_{s}\kappa \equiv \Delta_{c}\gamma_{\perp}-\Delta_{s}(\kappa_{g}-\kappa_{c})=0$, which gives the required gain rate $\kappa_{g}^{(0)} = \kappa_{c}+\gamma_{\perp}\Delta_{c}/\Delta_{s}$. The parameter-matching condition %$\Omega_{a}/\Omega_{J}=(\Delta_{c}-i\kappa)/\lambda$
is reduced to $\Omega_{a}/\Omega_{J}=\Delta_{c}(1+i\gamma_{\perp}/\Delta_{s})/\lambda$
and Eq.~(\ref{spin-1}) becomes
\begin{equation}\label{spin-2}
\begin{split}
&\bigg[1+\frac{\lambda^{2}}{\Delta_{c}\Delta_{s}(1+\gamma_{\perp}^{2}/\Delta_{s}^{2})}
       \frac{\langle J_{z}\rangle}{(N/2)}\bigg]\langle J_{-}\rangle=0,\\
&\frac{\gamma_{\perp}\lambda^{2}}{\Delta_{c}\Delta_{s}(1+\gamma_{\perp}^{2}/\Delta_{s}^{2})}
                                     \frac{\langle J_{+}\rangle \langle J_{-}\rangle}{(N/2)^{2}}
                                               -\gamma_{\parallel}\bigg(1+\frac{\langle J_{z}\rangle}{(N/2)}\bigg)=0.
\end{split}
\end{equation}
Now, besides the set of trivial solutions $\langle J_{z}\rangle/(N/2)=-1$ and $\langle J_{-}\rangle=0$, Eq.~(\ref{spin-2}) has also another set of {\it nontrivial} solutions
\begin{equation}\label{Jm}
 \frac{\langle J_{z}\rangle}{(N/2)}=-\frac{\lambda_{c}^{2}}{\lambda^{2}},~~~~~~
\frac{\langle J_{-}\rangle}{(N/2)}=\frac{\lambda_{c}}{\lambda}
              \bigg[\bigg(1-\frac{\lambda_{c}^{2}}{\lambda^{2}}\bigg)\frac{\gamma_{\parallel}}{\gamma_{\perp}}\bigg]^{1/2},
\end{equation}
with $\lambda_{c}$ modified as
\begin{equation}
\lambda_{c} \equiv \sqrt{\Delta_{c}\Delta_{s}(1+\gamma_{\perp}^{2}/\Delta_{s}^2)}.
\end{equation}
Here $\langle J_{-}\rangle$ is assumed to be real for simplicity. As shown in Fig.~\ref{figure2}, when varying the critical coupling strength $\lambda_{c}$ from $\lambda < \lambda_{c}$ to $\lambda > \lambda_{c}$, the proposed driven system exhibits a QPT from the normal phase with $\langle J_{z}\rangle/(N/2)=-1$ and $\langle J_{-}\rangle=0$ to the super-radiant phase with $\langle J_{z}\rangle$ and $\langle J_{-}\rangle$ given in Eq.~(\ref{Jm}). In the non-Hermitian case with gain, $\langle J_{z}\rangle/(N/2)$ has the same behavior as in the Hermitian case [see Fig.~\ref{figure2}(a)]. In fact, around $\lambda=\lambda_c$, $\langle J_{z}\rangle \sim |\lambda-\lambda_{c}|^{\nu_{z}}$, with $\nu_{z}=1$. Figure~\ref{figure2}(b) shows that in the non-Hermitian case with gain, $\langle J_{-}\rangle/(N/2)$ is much reduced in the regime of super-radiant phase, but it can be analytically derived from Eq.~(\ref{Jm}) that around $\lambda=\lambda_c$, $\langle J_{-}\rangle/(N/2)$ still exhibits the same critical behavior as in the Hermitian case: $\langle J_{-}\rangle \sim |\lambda-\lambda_{c}|^{\nu_{-}}$, with $\nu_{-}=1/2$.

\begin{figure}[tbp]
\centering
\includegraphics[width=0.48\textwidth]{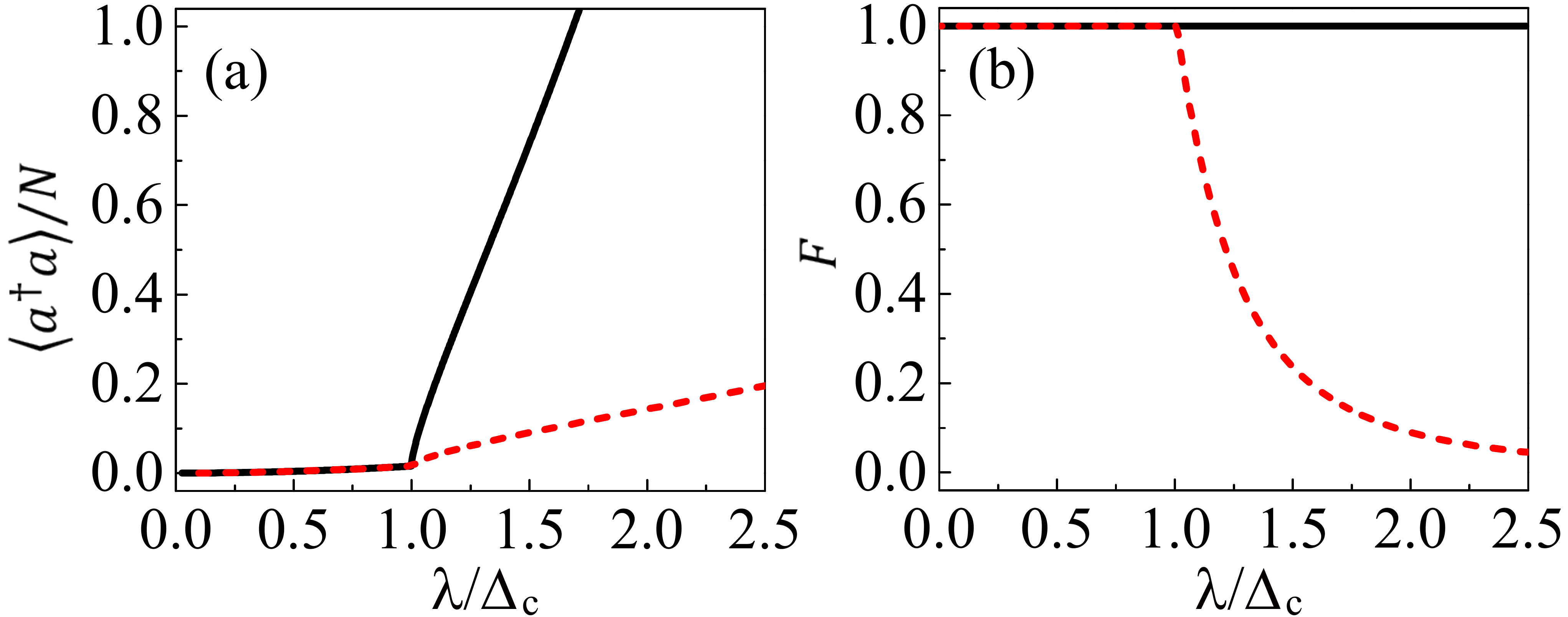}
\caption{(a) Mean photon number $\langle a^{\dag}a\rangle/N$ and (b) characteristic quantity $F$ of the TLS ensemble versus the reduced coupling strength $\lambda/\Delta_{c}$, where $\kappa_{c}=\gamma_{\perp}=\gamma_{\parallel}=\kappa_{g}=0$ for the (black) solid curve, and $\kappa_{c}=\gamma_{\perp}=1$, $\gamma_{\parallel}=0.1$ and $\kappa_{g}=\kappa_{g}^{(0)} = \kappa_{c}+\gamma_{\perp}\Delta_{c}/\Delta_{s}$ for the (red) dashed curve. Other parameters are the same as in Fig.~\ref{figure2}.}
\label{figure3}
\end{figure}

From the first equation in Eq.~(\ref{steady}), but with $\kappa_{c}$ replaced by $\kappa = \kappa_{c}-\kappa_{g}^{(0)} = -\gamma_{\perp}\Delta_{c}/\Delta_{s}$, we have
\begin{equation}\label{a}
\langle a\rangle=-\frac{\lambda}{\sqrt{N}\Delta_{c}(1+i\gamma_{\perp}/\Delta_{s})}\langle J_{-}\rangle
                   -\frac{\Omega_{a}}{\Delta_{c}(1+i\gamma_{\perp}/\Delta_{s})}.
\end{equation}
With $\gamma_{\perp}=0$, it reduces to the result in the ideal Hermitian case.
In Fig.~\ref{figure3}(a), we plot $\langle a^{\dag}a\rangle/N$ versus the reduced coupling strength $\lambda/\Delta_{c}$ in both the ideal Hermitian case and the non-Hermitian case with gain. In these two cases, the results for $\langle a^{\dag}a\rangle/N$  globally look different, but we can analytically derive that $\langle a^{\dag}a\rangle/N$ in the non-Hermitian case with gain also exhibits the same critical behavior as in the Hermitian case, i.e., around $\lambda=\lambda_c$, $\langle a^{\dag}a\rangle/N \sim |\lambda-\lambda_{c}|^{\nu_{a}}$, with $\nu_{a}=1/2$ (see Appendix \ref{appendix-A}).

Without decoherence in the system ($\kappa_c=\gamma_{\perp}=\gamma_{\parallel}=0$), $\Omega_{a}/\Omega_{J}=(\Delta_{c}-i\kappa_{c})/\lambda$
is reduced to the parameter-matching condition in the ideal Hermitian case, $\Omega_{a}/\Omega_{J}=\Delta_{c}/\lambda$, and the second equation in Eq.~(\ref{spin-1}) becomes trivial. In this Hermitian case, the results in Eqs.~(\ref{order-Jz}) and (\ref{order-J-}) are obtained by solving the first equation in Eq.~(\ref{spin-1}) and using the spin-conservation law $\langle J_{x}\rangle^2+\langle J_{y}\rangle^2+\langle J_{z}\rangle^2=(N/2)^{2}$~\cite{Soriente18,Zhu20}. In contrast, when the decoherence of the system is considered, the spin-conservation law is usually {\it violated}, with $\langle J_{x}\rangle^2+\langle J_{y}\rangle^2+\langle J_{z}\rangle^2 < (N/2)^{2}$~\cite{Torre16,Kirton17}. To characterize this violation, we define a characteristic quantity
$F=(\langle J_{x}\rangle^{2}+\langle J_{y}\rangle^{2}+\langle J_{z}\rangle^{2})/(N/2)^{2}$.
Obviously, $F=1$ in the ideal Hermitian case owing to the spin-conservation law. For the non-Hermitian case with gain, it is easy to verify that $F=1$ in the regime of normal phase, but
\begin{equation}\label{}
F=
\frac{\lambda_{c}^{2}}{\lambda^{2}}\bigg(1-\frac{\lambda_{c}^{2}}{\lambda^{2}}\bigg)\frac{\gamma_{\parallel}}{\gamma_{\perp}}
+\frac{\lambda_{c}^{4}}{\lambda^{4}}
\end{equation}
in the regime of super-radiant phase, where the ratio $\gamma_{\parallel}/\gamma_{\perp}$ between the longitudinal- and transversal-relaxation rates plays a significant role. In the non-Hermitian case with gain, $F$  monotonically decreases when $\lambda/\Delta_{c}>1.008$ (corresponding to $\lambda > \lambda_{c}$) [see the red dashed curve in Fig.~\ref{figure3}(b)].

\begin{figure}[tbp]
\centering
\includegraphics[width=0.48\textwidth]{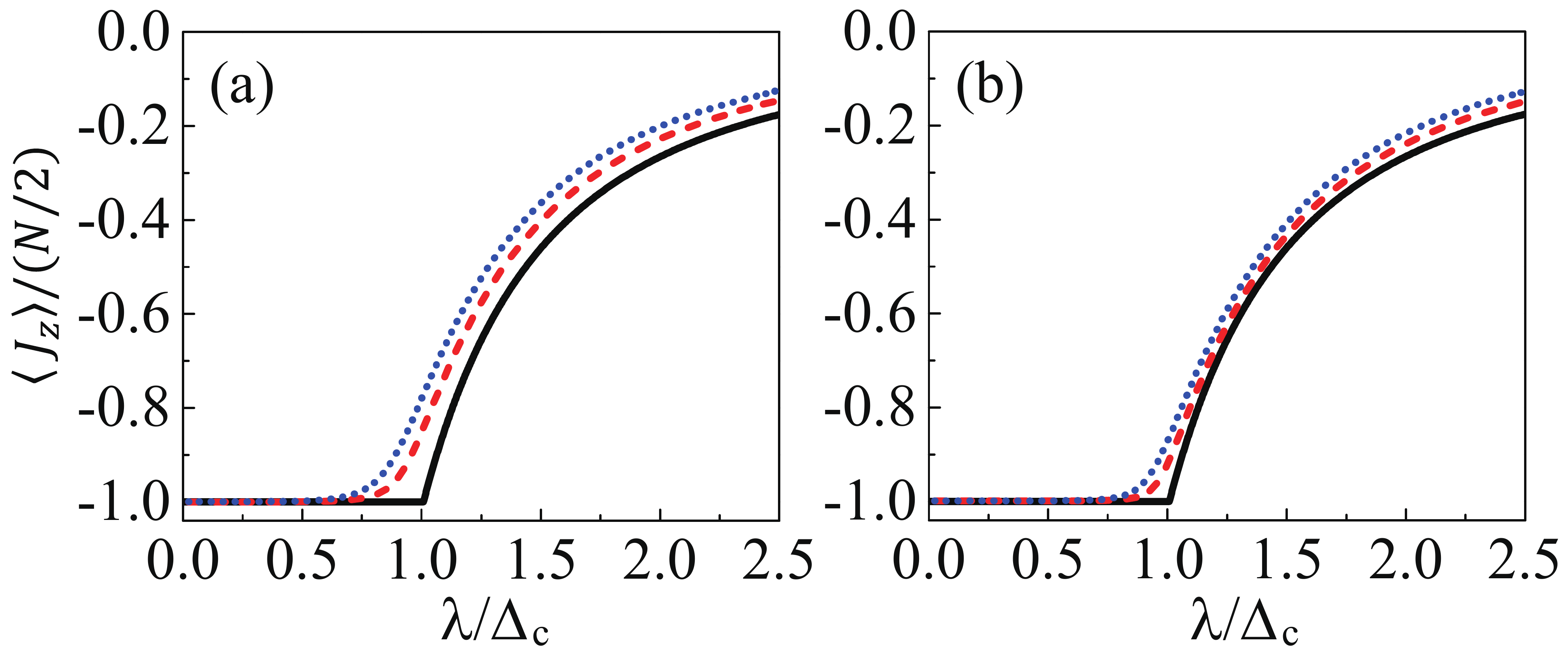}
\caption{(a) $\langle J_{z}\rangle/(N/2)$ versus the reduced coupling strength $\lambda/\Delta_{c}$ for $\kappa_{g}=\kappa_{g}^{(0)}$, where $\Omega_{a}/\Omega_{J}=\Delta_{c}(1+i\gamma_{\perp}/\Delta_{s})/\lambda$ for the (black) solid curve, $\Omega_{a}/\Omega_{J}=1.1\Delta_{c}(1+i\gamma_{\perp}/\Delta_{s})/\lambda$ for the (red) dashed curve, and $\Omega_{a}/\Omega_{J}=1.2\Delta_{c}(1+i\gamma_{\perp}/\Delta_{s})/\lambda$ for the (blue) dotted curve. (b) $\langle J_{z}\rangle/(N/2)$ versus the reduced coupling strength $\lambda/\Delta_{c}$ for $\Omega_{a}/\Omega_{J}=\Delta_{c}(1+i\gamma_{\perp}/\Delta_{s})/\lambda$, where $\kappa_{g}=\kappa_{g}^{(0)}$ for the (black) solid curve, $\kappa_{g}=1.2\kappa_{g}^{(0)}$ for the (red) dashed curve, and $\kappa_{g}=1.4\kappa_{g}^{(0)}$ for the (blue) dotted curve.  Other parameters are the same as in Fig.~\ref{figure2}.}
\label{figure4}
\end{figure}

So far we have systematically studied the QPT of the driven hybrid system in the parameter-matching case. However,
in the experiment, it is difficult to perfectly satisfy either the parameter-matching condition for the two drive fields or the condition for balancing the loss and gain in the system. Thus, it is useful to investigate the effect of the parameter mismatches on the QPT. In Fig.~\ref{figure4}, we plot $\langle J_{z}\rangle/(N/2)$ versus the reduced coupling strength $\lambda/\Delta_{c}$ in both cases of the drive-field mismatch and the gain-loss mismatch, as calculated from Eq.~(\ref{steady}) by replacing $\kappa_{c}$ with $\kappa = \kappa_{c}-\kappa_{g}$. For comparison, we also show the results for the perfect parameter-matching case (the black solid curve) in which both $\Omega_{a}/\Omega_{J}=\Delta_{c}(1+i\gamma_{\perp}/\Delta_{s})/\lambda$ and $\kappa_{g}=\kappa_{g}^{(0)}$ are satisfied. When $\kappa_{g}=\kappa_{g}^{(0)}$ but the two drive fields have a mismatch, i.e., $\Omega_{a}/\Omega_{J} \neq \Delta_{c}(1+i\gamma_{\perp}/\Delta_{s})/\lambda$, the cusp-like behavior of the QPT in the ideal case becomes obscured, as shown in the red dashed and blue dotted curves in Fig.~\ref{figure4}(a). On the other hand, when the two drive fields obey the condition $\Omega_{a}/\Omega_{J}=\Delta_{c}(1+i\gamma_{\perp}/\Delta_{s})/\lambda$ but $\kappa_{g} \neq \kappa_{g}^{(0)}$, the cusp-like behavior of the QPT in the ideal case becomes obscured as well [see the red dashed and blue dotted curves in Fig.~\ref{figure4}(b)]. Moreover, Eq.~(\ref{steady}) can have bistable solutions in the regime where the parameter-matching condition for the two drive fields is largely deviated (see Appendix \ref{appendix-D}). Here, it is not the case we focus on.

\subsection{Possible implementation}

In the experiment, the proposed model with gain can be implemented using a hybrid circuit-QED system composed of a dissipative TLS ensemble, such as the nitrogen-vacancy (NV) centers in diamond, coupled to an active coplanar-waveguide resonator [see Fig.~\ref{figure1}(b)]. The transition frequency of NV centers can be tuned by an external magnetic field and the number of NV centers in the sample can be $N \sim 10^{12}$~\cite{Kubo10,Amsuss11}, approaching the thermodynamic limit of the proposed system. The hybrid system is usually in the strong-coupling regime and can be described by a standard TC Hamiltonian in Eq.~(\ref{TC}).

To engineer an active resonator, one can harness an auxiliary flux qubit transversely coupled to the resonator with a coupling strength $g_{a}$ and longitudinally driven by two microwave fields~\cite{Quijandria18}. After eliminating the degree of freedom of the auxiliary qubit, the effective gain rate $\kappa_{g}$ of the resonator can be tuned from 0 to, e.g., $0.2g_{a}$ (i.e., $2\pi\times 6 $~MHz for $g_{a}/2\pi=30$~MHz) by varying the amplitudes and frequencies of the two microwave fields~\cite{Quijandria18}. Moreover, as shown in Fig.~\ref{figure1}(b), two additional drive fields with Rabi frequencies $\Omega_{a}$ and $\Omega_{J}$ pump the NV-center ensemble and the resonator, respectively. When the parameter-matching condition $\Omega_{a}/\Omega_{J}=\Delta_{c}(1+i\gamma_{\perp}/\Delta_{s})/\lambda$ is achieved for these two drive fields, an experimentally accessible TC model is then engineered and the relevant quantities for demonstrating the QPT are governed by Eqs.~(\ref{spin-2}) and (\ref{a}).

\section{Discussions and Conclusions}

The special case with only one drive field was studied in Refs.~\cite{Zou13,Emary04}, which is related to the cavity or the TLS ensemble pumped by a drive field, i.e., Eq.~(\ref{TC-d}) with $\Omega_{a}\neq 0$ but $\Omega_{J}=0$~\cite{Zou13} or $\Omega_{J} \neq 0$ but $\Omega_{a} = 0$~\cite{Emary04}. At a finite strength of this drive field, the biased TC model does not preserve the parity symmetry due to the bias term and the QPT disappears in the system~\cite{Emary04}. Only in the weak drive-field limit [i.e., either $\Omega_{a} \rightarrow 0$ when $\Omega_{J}=0$ or $\Omega_{J} \rightarrow  0$ when $\Omega_{a} = 0$ in Eq.~(\ref{TC-d})] can the model tend to have the QPT~\cite{Zou13}. For a realistic system, the cusp-like QPT behavior is however much smoothened by the inevitable cavity loss and the dissipation of the TLS ensemble even in the weak drive-field limit~\cite{Feng15}. In our proposal, when the two drive fields satisfy the parameter-matching condition, we can obtain an effective TC Hamiltonian with parity symmetry [i.e., Eq.~(\ref{eff})]. This TC Hamiltonian can exhibit QPT at an experimentally-accessible critical coupling strength and the weak drive-field limit is not necessary. Even with the decoherence of the system, the QPT behavior is still achievable by harnessing an active cavity (instead of a passive cavity). Very recently, the QPT in a TC model induced by a single {\it squeezed} drive field was investigated~\cite{Zhu20}, where the effective Hamiltonian is a biased TC model with a two-photon drive term, corresponding to Eq.~(\ref{TC-d}) with $\Omega_{a}a^{\dag}+\Omega_{a}^{*}a$ replaced by $\Omega_{a}a^{{\dag}2}+\Omega_{a}^{*}a^{2}$ and $\Omega_{J}=0$. Using a squeezing transformation, this biased TC Hamiltonian can be transformed to an anisotropic Dicke Hamiltonian~\cite{Qin18}. Experimentally, if a {\it pure} squeezed drive field, $a^{{\dag}2}+a^{2}$, cannot be achieved, as discussed above, any finite {\it unsqueezed} part of the single drive field can ruin the QPT. In our study, the obtained effective Hamiltonian in Eq.~(\ref{eff}) is a standard TC model without the bias term and no squeezed drive field is required.
Also, the spin-conservation law is used in Ref.~\cite{Zhu20}. It is different from our non-Hermitian case in which the spin-conservation law is found to be violated~\cite{Torre16,Kirton17}.

In conclusion, we have studied the QPT in a TC model engineered with two drive fields applied to the TLS ensemble and the cavity, respectively. In the ideal Hermitian case without decoherence, the QPT can occur at an experimentally-accessible critical coupling strength, but it is spoiled by the decoherence of the system. In this non-Hermitian case, we find that the QPT can be recovered by harnessing a gain in the cavity to balance the loss of the TLS ensemble. In sharp contrast to the spin conservation in the ideal Hermitian case, the spin-conservation law is however found to be violated in our non-Hermitian case. Moreover, we propose to implement this non-Hermitian TC model using a hybrid circuit-QED system. Our work provides an experimentally realizable approach to achieving QPT in the non-Hermitian TC model.

\section*{Acknowledgments}
This work is supported by the National Key Research and Development Program of China (Grant No.~2016YFA0301200), the National Natural Science Foundation of China (Grants No.~11934010, No.~U1801661, and No.~11774022), the Postdoctoral Science Foundation of China (Grant No.~2020M671687), and Zhejiang Province Program for Science and Technology (Grant No.~2020C01019).

\appendix

\section{Critical exponent of the mean photon number in the cavity}\label{appendix-A}

\subsection{The ideal Hermitian case}

In the ideal Hermitian case without decoherence in the system, it follows from Eq.~(\ref{a}) that
\begin{equation}\label{}
\langle a\rangle=-\frac{\lambda}{\sqrt{N}\Delta_{c}}\langle J_{-}\rangle
                   -\frac{\Omega_{a}}{\Delta_{c}},
\end{equation}
by having $\gamma_{\perp}=0$, where $\langle J_{-}\rangle$ is given by Eq.~(\ref{order-J-}). The mean photon number $\langle a^{\dag}a\rangle$ in the cavity can be written as
\begin{equation}\label{}
\begin{split}
\frac{\langle a^{\dag}a\rangle}{N}
=\Bigg\{
\begin{array}{cc}
           \Big|\frac{(\Omega_{a}/\sqrt{N})}{\Delta_{c}}\Big|^{2},                         &~~~ \lambda < \lambda_{c};\\
   \Big|\frac{\lambda}{2\Delta_{c}}\Big(1-\frac{\lambda_{c}^{4}}{\lambda^{4}}\Big)^{1/2}+
       \frac{(\Omega_{a}/\sqrt{N})}{\Delta_{c}} \Big|^{2},    &~~~ \lambda \geq \lambda_{c}.
\end{array}
\end{split}
\end{equation}
To reveal the critical behavior of the mean photon number, we study the variation of the mean photon number around the critical coupling strength $\lambda_{c}=\sqrt{\Delta_{c}\Delta_{s}}$,
\begin{eqnarray}\label{variation}
\delta n &\equiv& \frac{\langle a^{\dag}a\rangle}{N}-\frac{\langle a^{\dag}a\rangle}{N}\bigg|_{\lambda=\lambda_{c}} \nonumber\\
&=&\frac{\lambda^{2}}{4\Delta_{c}^{2}}\Big(1-\frac{\lambda_{c}^{4}}{\lambda^{4}}\Big)+
  \frac{\lambda(\Omega_{a}/\sqrt{N})}{\Delta_{c}^{2}}\Big(1-\frac{\lambda_{c}^{4}}{\lambda^{4}}\Big)^{1/2}\\
& &      +\bigg[\frac{(\Omega_{a}/\sqrt{N})^{2}}{\Delta_{c}^{2}}
       -\frac{(\Omega_{a}/\sqrt{N})^{2}}{\Delta_{c}^{2}}\bigg|_{\lambda=\lambda_{c}}\bigg].\nonumber
\end{eqnarray}
Here we assume the Rabi frequency $\Omega_{a}$ to be real. In our model, a finite drive power ($\Omega_{a}\neq 0$) is applied to the cavity, so the critical behavior of the mean photon number around the critical coupling strength $\lambda_{c}$ is dominantly determined by the second term in Eq.~(\ref{variation}). Therefore, we have
\begin{equation}\label{}
\lim_{\lambda \rightarrow \lambda_{c}}\delta n
=\frac{2(\Omega_{a}/\sqrt{N})\sqrt{\lambda_{c}}}{\Delta_{c}^{2}}(\lambda-\lambda_{c})^{1/2}
                                               \sim (\lambda-\lambda_{c})^{\nu_{a}},
\end{equation}
with the critical exponent $\nu_{a}=1/2$.

\subsection{The non-Hermitian case with gain}

In the non-Hermitian case with gain, it follows from Eqs.~(\ref{Jm}) and (\ref{a}) that the mean photon number $\langle a^{\dag}a\rangle$ in the cavity can be written as
\begin{equation}\label{}
\begin{split}
\frac{\langle a^{\dag}a\rangle}{N}
=\Bigg\{
\begin{array}{cc}
           \Big|\frac{(\Omega_{a}/\sqrt{N})}{\Delta_{c}(1+i\gamma_{\perp}/\Delta_{s})}\Big|^{2}, & \lambda < \lambda_{c};\\
   \Big|\frac{\lambda_{c}}{2\Delta_{c}(1+i\gamma_{\perp}/\Delta_{s})}
              \Big[\Big(1-\frac{\lambda_{c}^{2}}{\lambda^{2}}\Big)\frac{\gamma_{\parallel}}{\gamma_{\perp}}\Big]^{1/2}
              +\frac{(\Omega_{a}/\sqrt{N})}{\Delta_{c}(1+i\gamma_{\perp}/\Delta_{s})} \Big|^{2}, & \lambda \geq \lambda_{c},
\end{array}
\end{split}
\end{equation}
where the critical coupling strength is modified as $\lambda_{c} \equiv \sqrt{\Delta_{c}\Delta_{s}(1+\gamma_{\perp}^{2}/\Delta_{s}^2)}$.
The variation of the mean photon number around the critical coupling strength $\lambda_{c}$ is
\begin{eqnarray}\label{variation2}
\delta n &\equiv& \frac{\langle a^{\dag}a\rangle}{N}-\frac{\langle a^{\dag}a\rangle}{N}\bigg|_{\lambda=\lambda_{c}} \nonumber\\
&=&\frac{\lambda_{c}^{2}}{4\Delta_{c}^{2}(1+\gamma_{\perp}^{2}/\Delta_{s}^{2})}
              \Big[\Big(1-\frac{\lambda_{c}^{2}}{\lambda^{2}}\Big)\frac{\gamma_{\parallel}}{\gamma_{\perp}}\Big]\nonumber\\
& &            +\frac{\lambda_{c}(\Omega_{a}/\sqrt{N})}{\Delta_{c}^{2}(1+\gamma_{\perp}^{2}/\Delta_{s}^{2})}
              \Big[\Big(1-\frac{\lambda_{c}^{2}}{\lambda^{2}}\Big)\frac{\gamma_{\parallel}}{\gamma_{\perp}}\Big]^{1/2}\\
& &            +\bigg[\frac{(\Omega_{a}/\sqrt{N})^{2}}{\Delta_{c}^{2}(1+\gamma_{\perp}^{2}/\Delta_{s}^{2})}
          -\frac{(\Omega_{a}/\sqrt{N})^{2}}{\Delta_{c}^{2}(1+\gamma_{\perp}^{2}/\Delta_{s}^{2})}
                  \bigg|_{\lambda=\lambda_{c}}\bigg].\nonumber
\end{eqnarray}
The critical behavior of the mean photon number around the critical coupling strength $\lambda_{c}$ is also dominantly determined by the second term in Eq.~(\ref{variation2}). Then, we obtain the critical behavior of $\langle a^{\dag}a\rangle$ in the considered non-Hermitian case with gain,
\begin{eqnarray}\label{}
\lim_{\lambda \rightarrow \lambda_{c}}\delta n =
              \frac{(\Omega_{a}/\sqrt{N})\sqrt{2\lambda_{c}\gamma_{\parallel}/\gamma_{\perp}}}
                {\Delta_{c}^{2}(1+\gamma_{\perp}^{2}/\Delta_{s}^{2})}(\lambda-\lambda_{c})^{1/2}
                                               \sim (\lambda-\lambda_{c})^{\nu_{a}},~~~~~~~
\end{eqnarray}
which has the same critical exponent $\nu_{a}=1/2$ as in the Hermitian case.

\section{Derivation of Eq.~(\ref{mean}) via a master equation approach}\label{appendix-B}

In the non-Hermitian case, we can also study the QPT using a master equation approach. With the relations $J_{\alpha}=\frac{1}{2}\sum_{k=1}^{N}\sigma_{\alpha}^{(k)}$ and $J_{\pm}=\sum_{k=1}^{N}\sigma_{\pm}^{(k)}$, $\alpha=x$, $y$, $z$, the Hamiltonian of the driven system in Eq.~(\ref{TC-d}) can be rewritten as
\begin{eqnarray}\label{TC-TLS}
H^{(d)}_{\rm TC}&=&\Delta_{c} a^{\dag}a + \frac{\Delta_{s}}{2}\sum_{k=1}^{N}\sigma_{z}^{(k)}
                   +\frac{\lambda}{\sqrt{N}}\sum_{k=1}^{N}[a\sigma_{+}^{(k)}+a^{\dag}\sigma_{-}^{(k)}]\nonumber\\
                & & +(\Omega_{a}a^{\dag}+\Omega_{a}^{*}a)
                    +\frac{1}{\sqrt{N}}\sum_{k=1}^{N}[\Omega_{J}\sigma_{+}^{(k)}+\Omega_{J}^{*}\sigma_{-}^{(k)}],
\end{eqnarray}
where $\sigma_{\alpha}^{(k)}$ are the spin-$1/2$ Pauli operators related to the $k$th TLS in the ensemble, and $\sigma_{\pm}^{(k)}=[\sigma_{x}^{(k)}\pm i\sigma_{y}^{(k)}]/2$ are the corresponding raising and lowering operators.

The master equation for the density matrix $\rho$ of the system can be written, in the Lindblad form, as~\cite{Breuer07}
\begin{eqnarray}\label{master}
\dot{\rho}&=&i[\rho,H]+\kappa_{c}(2a\rho a^{\dag}-a^{\dag}a\rho -\rho a^{\dag}a)
             +\gamma_{p}\sum_{k=1}^{N}[\sigma_{z}^{(k)}\rho \sigma_{z}^{(k)}-\rho]\nonumber\\
          & &+\gamma_{h}\sum_{k=1}^{N}[2\sigma_{-}^{(k)}\rho \sigma_{+}^{(k)}-\sigma_{+}^{(k)}\sigma_{-}^{(k)}\rho
                                                                             -\rho \sigma_{+}^{(k)}\sigma_{-}^{(k)}],
\end{eqnarray}
where $\gamma_{p}$ and $\gamma_{h}$ are the {\it nonradiative} dephasing rate and the {\it radiative} decay rate of the individual TLSs, respectively, while $\kappa_c$ is the decay rate of the cavity. From the master equation in Eq.~(\ref{master}), we can derive the equation of motion for the expectation value of the operator $\mathcal{O}$ $(=\{a,\sigma_{-}^{(k)},\sigma_{z}^{(k)}\})$ via the relation $\langle\dot{\mathcal{O}}\rangle=\rm{Tr}[\dot{\rho}\mathcal{O}]$,
\begin{eqnarray}\label{spin-b1}
\langle\dot{a}\rangle
&=&-i(\Delta_{c}-i\kappa_{c})\langle a\rangle
  -i\frac{\lambda}{\sqrt{N}}\sum_{k=1}^{N}\langle \sigma_{-}^{(k)} \rangle-i\Omega_{a},\nonumber\\
\langle\dot{\sigma}_{-}^{(k)}\rangle
&=&-i(\Delta_{s}-i\gamma_{\perp})\langle\sigma_{-}^{(k)}\rangle
  +i\frac{\lambda}{\sqrt{N}}\langle\sigma_{z}^{(k)} a \rangle
  +i\langle\sigma_{z}^{(k)}\rangle\frac{\Omega_{J}}{\sqrt{N}},\nonumber\\
\langle\dot{\sigma}_{z}^{(k)}\rangle
&=& -\gamma_{\parallel}(1+\langle \sigma_{z}^{(k)}\rangle)
  -2i\frac{\lambda}{\sqrt{N}}(\langle a\sigma_{+}^{(k)}\rangle-\langle a^{\dag}\sigma_{-}^{(k)}\rangle)\\
& &-2i\frac{1}{\sqrt{N}}(\Omega_{J}\langle \sigma_{+}^{(k)}\rangle-\Omega_{J}^{*}\langle \sigma_{-}^{(k)}\rangle),
     \nonumber
\end{eqnarray}
where the transverse relaxation rate is $\gamma_{\perp}=\gamma_{h}+2\gamma_{p}$ and the longitudinal relaxation rate is $\gamma_{\parallel}=2\gamma_{h}$. Summing the second and third equations in Eq.~(\ref{spin-b1}) over all TLSs in the ensemble, respectively, we obtain
\begin{eqnarray}\label{spin-b3}
\langle\dot{a}\rangle
&=&-i(\Delta_{c}-i\kappa_{c})\langle a\rangle
  -i\frac{\lambda}{\sqrt{N}}\langle J_{-} \rangle-i\Omega_{a},\nonumber\\
\langle\dot{J}_{-}\rangle
&=&-i(\Delta_{s}-i\gamma_{\perp})\langle J_{-}\rangle
  +i\frac{2\lambda}{\sqrt{N}}\langle J_{z} a \rangle+i2\langle J_{z}\rangle\frac{\Omega_{J}}{\sqrt{N}},\\
\langle\dot{J}_{z}\rangle
&=&-i\frac{\lambda}{\sqrt{N}}(\langle aJ_{+}\rangle-\langle a^{\dag}J_{-}\rangle)
  -i\frac{1}{\sqrt{N}}(\Omega_{J}\langle J_{+}\rangle-\Omega_{J}^{*}\langle J_{-}\rangle)\nonumber\\
& & -\gamma_{\parallel}\bigg(\frac{N}{2}+\langle J_{z}\rangle\bigg),\nonumber
\end{eqnarray}
where the relations $J_{z}=\frac{1}{2}\sum_{k=1}^{N}\sigma_{z}^{(k)}$ and $J_{\pm}=\sum_{k=1}^{N}\sigma_{\pm}^{(k)}$ are used. This is just Eq.~(\ref{mean}) in the main text.

\section{Results beyond the mean-field approximation in Eq.~(\ref{steady})}\label{appendix-C}

\begin{figure}[tbp]
\centering
\includegraphics[width=0.48\textwidth]{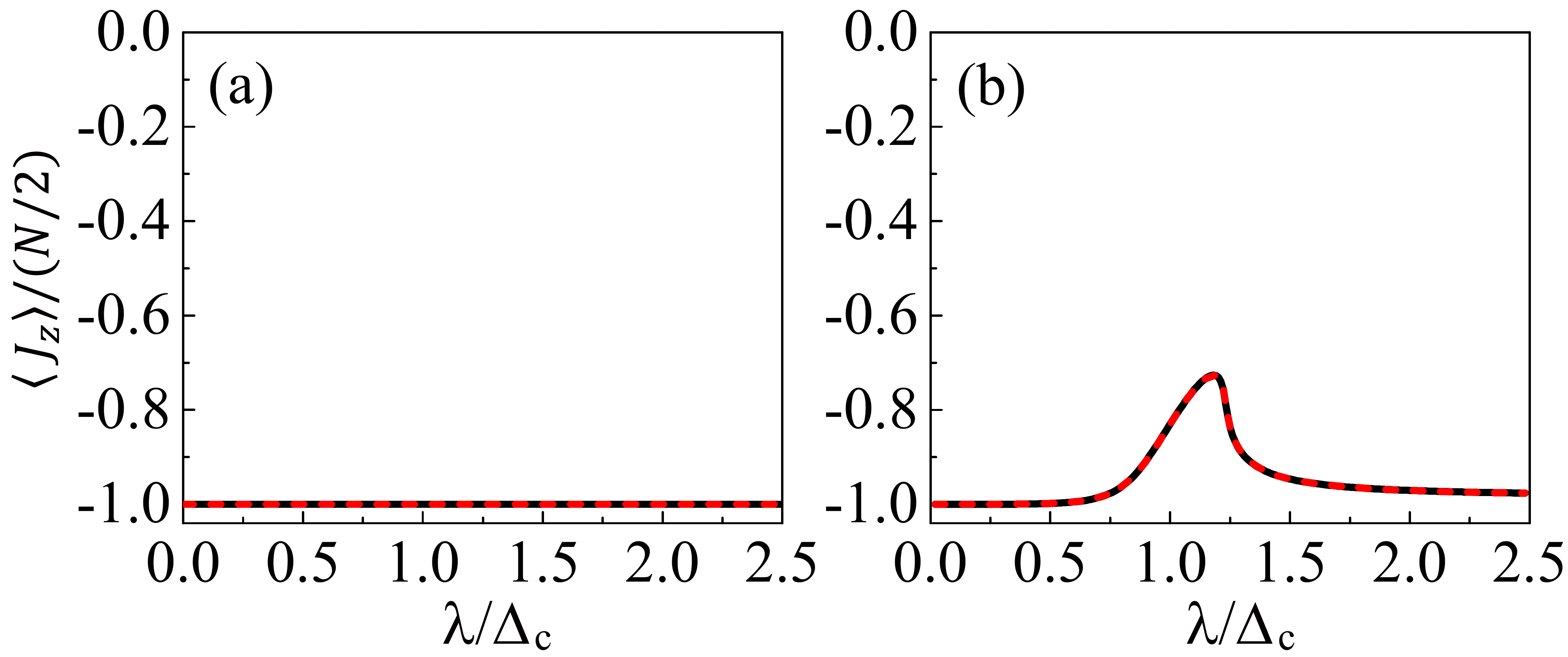}
\caption{$\langle J_{z}\rangle/(N/2)$ versus the reduced coupling strength $\lambda/\Delta_{c}$ under (a) the parameter-matching condition $\Omega_{a}/\Omega_{J}=(\Delta_{c}-i\kappa_{c})/\lambda$ and (b) the parameter-mismatching condition $\Omega_{a}/\Omega_{J}=1.2(\Delta_{c}-i\kappa_{c})/\lambda$ for the two drive fields, where the (black) solid curves are calculated using Eq.~(\ref{steady}), while the (red) dashed curves are calculated using both Eqs.~(\ref{c1}) and (\ref{c4}). Here $\kappa_{g}=0$, which corresponds to the passive-cavity case. Other parameters are the same as in Fig.~\ref{figure2}.}
\label{figure-s1}
\end{figure}

In Sec.~\ref{non-Hermitian}, we have used the mean-field approximation, $\langle J_{z}a\rangle=\langle J_{z}\rangle\langle a\rangle$ and $\langle aJ_{+}\rangle=\langle a\rangle\langle J_{+}\rangle$, and neglect the correlations due to the two-operator terms $J_{z}a$ and $aJ_{+}$. Below we consider these correlations. Without the mean-field approximation, Eq.~(\ref{mean}) at the steady state is reduced to
\begin{equation}\label{c1}
\begin{split}
&(\Delta_{c}-i\kappa_{c})\langle a\rangle+\frac{\lambda}{\sqrt{N}}\langle J_{-}\rangle+\Omega_{a}=0,\\
&(\Delta_{s}-i\gamma_{\perp})\langle J_{-}\rangle-\frac{2\lambda}{\sqrt{N}}\langle J_{z} a \rangle
                      -2\langle J_{z}\rangle\frac{\Omega_{J}}{\sqrt{N}}=0,\\
&\frac{\lambda}{\sqrt{N}}(\langle a J_{+}\rangle-\langle a^{\dag}J_{-}\rangle)+\frac{1}{\sqrt{N}}(\Omega_{J}\langle J_{+}\rangle-\Omega_{J}^{*}\langle J_{-}\rangle)\\
&                   -i\gamma_{\parallel}\bigg(\frac{N}{2}+\langle J_{z}\rangle\bigg)=0.
\end{split}
\end{equation}
Also, with the quantum Langevin approach, we can further write the equations of motion for the two-operator terms $J_{z}a$ and $aJ_{+}$,
\begin{eqnarray}\label{c2}
\begin{split}
\frac{d}{d t} (J_{z}a)=&-i\big[\Delta_{c}-i(\kappa_{c}+\gamma_{\|})\big]J_{z}a-\frac{N}{2}\gamma_{\|}a
                         -i\frac{\lambda}{\sqrt{N}}(J_{z}J_{-}\\
                       &-a^{\dag}aJ_{-}+aaJ_{+})
                        -i\frac{1}{\sqrt{N}}(\Omega_{J}aJ_{+}-\Omega_{J}^{*}aJ_{-})\\
                       & -i\Omega_{a}J_{z}+\sqrt{2\kappa_{c}}J_{z}a_{\rm in}
                           +\sqrt{2\gamma_{\parallel}}\,aJ_{\rm z,\,in},\\
\frac{d}{d t} (aJ_{+})=&-i\big[(\Delta_{c}-\Delta_{s})-i(\kappa_{c}+\gamma_{\perp})\big]aJ_{+}
                        -i\frac{\lambda}{\sqrt{N}}(J_{-}J_{+}\\
                       &+2J_{z}aa^{\dag})
                        -i\Omega_{a}J_{+}-i2\frac{\Omega_{J}^{*}}{\sqrt{N}}J_{z}a+\sqrt{2\kappa_{c}}J_{+}a_{\rm in}\\
                       & +\sqrt{2\gamma_{\perp}}aJ_{+,\,\rm in}.
\end{split}
\end{eqnarray}
Here we consider the leading contributions from $J_{z}a$ and $aJ_{+}$ by applying a mean-field approximation to other multiple-operator terms in Eq.~(\ref{c2}). Thus, we have
\begin{eqnarray}\label{c3}
\begin{split}
\frac{d}{d t} \langle J_{z}a\rangle=&-i\big[\Delta_{c}-i(\kappa_{c}+\gamma_{\|})\big]\langle J_{z}a\rangle
                                       -\frac{N}{2}\gamma_{\|}\langle a\rangle\\
                        & -i\frac{\lambda}{\sqrt{N}}(\langle J_{z}\rangle \langle J_{-}\rangle
                         -\langle a^{\dag}J_{-}\rangle\langle a \rangle + \langle aJ_{+}\rangle\langle a\rangle)\\
                        &-i\frac{1}{\sqrt{N}}(\Omega_{J}\langle aJ_{+}\rangle
                         -\Omega_{J}^{*}\langle a\rangle \langle J_{-}\rangle)-i\Omega_{a}\langle J_{z}\rangle,\\
\frac{d}{d t} \langle aJ_{+}\rangle=&-i\big[(\Delta_{c}-\Delta_{s})-i(\kappa_{c}+\gamma_{\perp})\big]\langle aJ_{+}\rangle
                        -i\frac{\lambda}{\sqrt{N}}(\langle J_{-}\rangle \langle J_{+}\rangle\\
                       &                      +2\langle J_{z}a\rangle \langle a^{\dag}\rangle)
                        -i\Omega_{a}\langle J_{+}\rangle-i2\frac{\Omega_{J}^{*}}{\sqrt{N}}\langle J_{z}a\rangle.
\end{split}
\end{eqnarray}
At the steady state, $d\langle J_{z}a\rangle/dt=d\langle aJ_{+}\rangle/dt=0$, which gives
\begin{eqnarray}\label{c4}
\begin{split}
&\big[\Delta_{c}-i(\kappa_{c}+\gamma_{\|})\big]\langle J_{z}a\rangle
                                       -i\frac{N}{2}\gamma_{\|}\langle a\rangle
                         +\frac{\lambda}{\sqrt{N}}(\langle J_{z}\rangle \langle J_{-}\rangle
                         -\langle a^{\dag}J_{-}\rangle\langle a \rangle\\
                       & + \langle aJ_{+}\rangle\langle a\rangle)
                         +\frac{1}{\sqrt{N}}(\Omega_{J}\langle aJ_{+}\rangle-\Omega_{J}^{*}\langle a\rangle \langle J_{-}\rangle)+\Omega_{a}\langle J_{z}\rangle=0,\\
&\big[(\Delta_{c}-\Delta_{s})-i(\kappa_{c}+\gamma_{\perp})\big]\langle aJ_{+}\rangle
                        +\frac{\lambda}{\sqrt{N}}(\langle J_{-}\rangle \langle J_{+}\rangle
                                             +2\langle J_{z}a\rangle \langle a^{\dag}\rangle)\\
                       &+\Omega_{a}\langle J_{+}\rangle+2\frac{\Omega_{J}^{*}}{\sqrt{N}}\langle J_{z}a\rangle=0.
\end{split}
\end{eqnarray}
We can also study the steady-state behaviors of the system by numerically solving both Eqs.~(\ref{c1}) and (\ref{c4}), which are beyond the mean-field approximation in Eq.~(\ref{steady}).

\begin{figure}[tbp]
\centering
\includegraphics[width=0.48\textwidth]{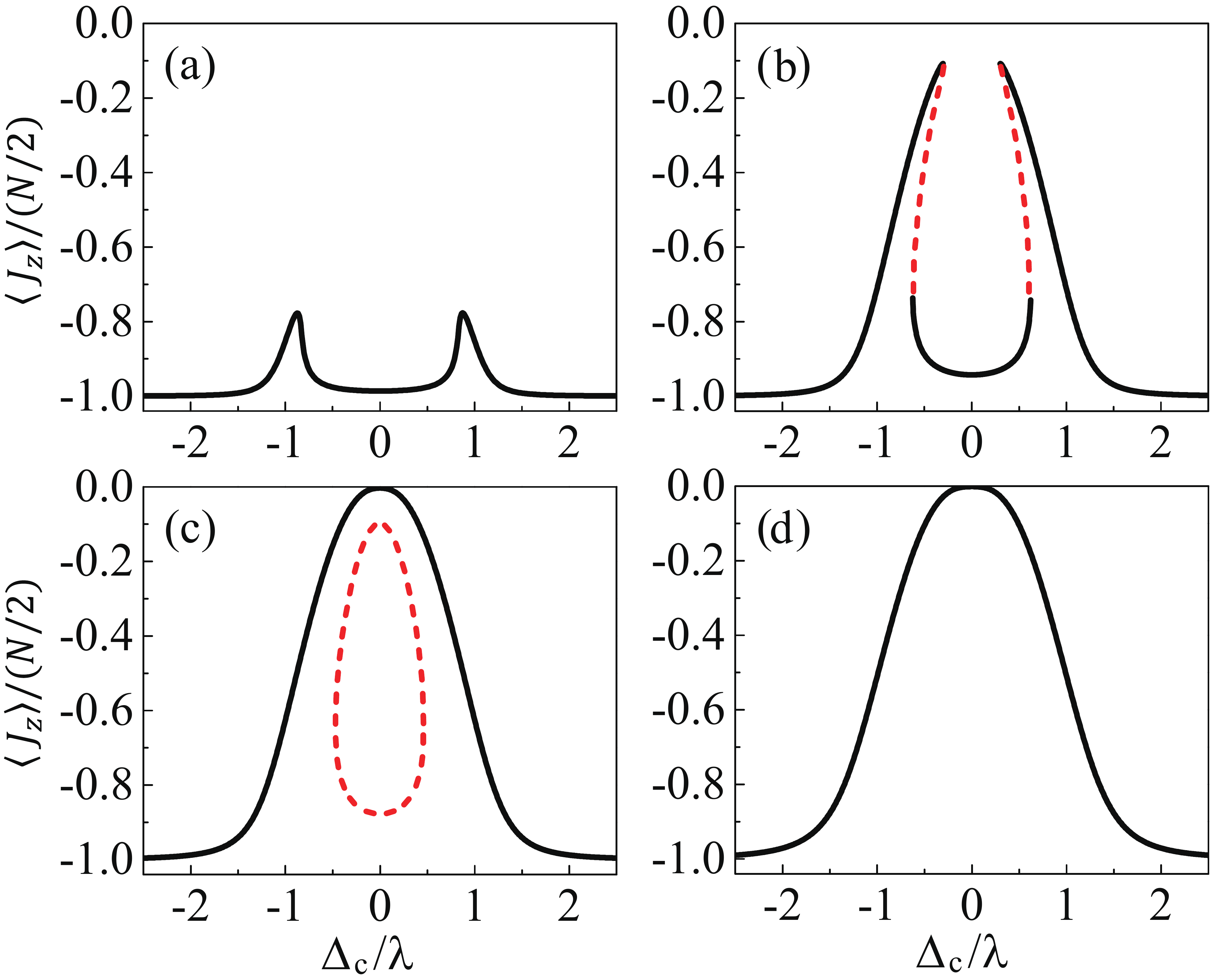}
\caption{$\langle J_{z}\rangle/(N/2)$ versus the reduced frequency detuning $\Delta_{c}/\lambda$ in the passive-cavity case of $\kappa_{g}=0$, calculated using Eq.~(\ref{steady}), where (a) $\Omega_{a}/\Omega_{J}=0.87(\Delta_{c}-i\kappa_{c})/\lambda$, (b) $\Omega_{a}/\Omega_{J}=0.77(\Delta_{c}-i\kappa_{c})/\lambda$, (c) $\Omega_{a}/\Omega_{J}=0.7(\Delta_{c}-i\kappa_{c})/\lambda$, and (d) $\Omega_{a}/\Omega_{J}=0.6(\Delta_{c}-i\kappa_{c})/\lambda$. Other parameters are the same as in Fig.~\ref{figure2}.}
\label{figure-s2}
\end{figure}

In Fig.~\ref{figure-s1}, we plot the expectation value $\langle J_{z}\rangle/(N/2)$ versus the reduced coupled strength $\lambda/\Delta_{c}$ under the parameter-matching condition $\Omega_{a}/\Omega_{J}=(\Delta_{c}-i\kappa_{c})/\lambda$ and the parameter-mismatching condition $\Omega_{a}/\Omega_{J}=1.2(\Delta_{c}-i\kappa_{c})/\lambda$ for the two drive fields, respectively (see the red dashed curves). These results are obtained by numerically solving both Eqs.~(\ref{c1}) and (\ref{c4}). Obviously, they are nearly identical to those obtained from Eq.~(\ref{steady}) (comparing the red dashed curves with the black solid curves in Fig.~\ref{figure-s1}). This clearly shows that the mean-field approximation in Eq.~(\ref{steady}) can indeed give accurate results in the thermodynamic limit of the TLS ensemble, as demonstrated in Ref.~\cite{Krimer19} as well.

\section{Bistability of the system}\label{appendix-D}

Below we show that when the deviation of the parameter-matching condition for the two drive fields becomes sufficiently large, bistability can occur in the driven system. Here we take the passive-cavity case of $\kappa_{g}=0$ as an example to demonstrate this phenomenon. In Fig.~\ref{figure-s2}, we plot $\langle J_{z}\rangle/(N/2)$ versus the reduced frequency detuning $\Delta_{c}/\lambda$ for various drive-field mismatches. When the two drive fields satisfy the parameter-matching condition $\Omega_{a}/\Omega_{J}=(\Delta_{c}-i\kappa_{c})/\lambda$, the effects of the two drive fields cancel each other, corresponding to a zero effective drive strength on the system. It is expected that this effective drive strength increases with the deviation of the parameter-matching condition for the two drive fields. For a small deviation (i.e., weak effective drive strength), the TLS ensemble is in low-lying excitations and the coupled TLS ensemble-cavity system can be approximated by a model of two coupled harmonic oscillators (see, e.g., Ref.~\cite{Zhang-npj}). In such a case, $\langle J_{z}\rangle/(N/2)$ is a single-valued function of the reduced frequency detuning $\Delta_{c}/\lambda$, which has two peaks around $\Delta_{c}/\lambda=\pm 1$ [see Fig.~\ref{figure-s2}(a)].

When the deviation is sufficiently large (corresponding to a strong effective drive strength), the coupled TLS ensemble-cavity system exhibits the bistable behavior [see Fig.~\ref{figure-s2}(b)], where the higher and lower branches are stable and the intermediate branch (the red dashed parts of the curve) is unstable. This is similar to the bistable phenomenon observed in Ref.~\cite{Angerer17}. When further increasing the deviation of the parameter-matching condition for the two drive fields, the parts of the curve corresponding to the unstable solutions form a closed loop [see the red dashed parts of the curve in Fig.~\ref{figure-s2}(c)]. For an even larger deviation of the parameter-matching condition for the two drive fields, $\langle J_{z}\rangle/(N/2)$ becomes a single-valued function of $\Delta_{c}/\lambda$ again [see Fig.~\ref{figure-s2}(d)]. Now, the TLS ensemble is in the state of very high excitations and it is approximately decoupled from the cavity. In the active-cavity case with gain, we can also obtain the similar results above. In the experiment, the parameter-matching condition for the two drive fields can be closely achieved. It is in the regime that has a small deviation of the parameter-matching condition for the two drive fields, where the bistability does not occur in the coupled TLS ensemble-cavity system.

\end{document}